# Frequency modulated laser beam with highly efficient intensity stabilisation

Frédéric du Burck, Assia Tabet, Olivier Lopez

Laboratoire de Physique des Lasers,

Université Paris 13 – CNRS UMR 7538,

99 avenue Jean-Baptiste Clément,

93430, Villetaneuse, France

duburck@galilee.univ-paris13.fr

**Abstract**

We analyse the limitation of the amplitude modulation rejection due to the spatial modulation of the output beam of an acousto-optic modulator used in an active laser beam stabilisation system when a frequency modulation of a few megahertz is applied to this modulator. We show how to overcome this problem, using a single mode optical fibre at the output of the modulator. A residual amplitude modulation of $10^{-5}$ is achieved.

In a previous work, we have presented an active system for the stabilisation of the beam intensity around a carrier frequency in the megahertz range, based on an adaptive narrow-band controller driving the RF input power of an acousto-optic modulator (AOM) [1]. We applied recently this system to the correction of the residual amplitude modulation (AM) of the laser beam in frequency modulation (FM) spectroscopy experiments carried out in iodine vapour in a cell [2]. In this technique, the probe beam is frequency or phase modulated at a frequency much greater than the width of the studied line. During the interaction with the sample, each Fourier component of the modulated beam experiences attenuation and phase shift. The beat notes between them on a photodetector result in photocurrent components at the modulation frequency and its harmonics [3,4]. The sensitivity of FM spectroscopy is limited by the AM generated by the modulator because a part of the low frequency technical noise of the laser is transferred to the detection frequency [5]. Moreover, it gives a non-zero baseline and a distortion of the detected line shape [6]. In order to simplify our scheme for FM spectroscopy, the FM signal is applied directly to the AOM used for the control of the beam. It appears that an important limitation of the AM rejection is due to the angular separation of the Fourier components of the modulated beam at the AOM output. In this letter, we analyse this limitation and we show how it can be overcome using a single mode optical fibre located at the AOM output. A frequency modulated beam in the MHz range with a $10^{-5}$ residual AM at the modulation frequency is then achieved.

Fig. 1 depicts the set-up for AM rejection. The beam is frequency shifted by 250 MHz and frequency modulated at $f_m = 2.5\,\text{MHz}$ by the AOM (AA SHT 250 A&A) with a modulation index near 1. The spectrum of the FM signal extends on about 10 MHz around 250 MHz. The fluctuations of the frequency response of the device in this frequency band result in a few % AM of the beam at the modulation frequency. After crossing a telescope with a pinhole, in order to limit the transit effect in the $I_2$ cell and to improve the wave fronts quality, the beam is split into beam 1, the correction beam whose intensity is monitored by photodiode PD1 to generate the error signal for the narrow-band controller, and beam 2, the useful beam. This latter interacts with the iodine vapour during spectroscopy experiments. Its intensity is measured by photodiode PD2. The cancelling signal generated by the narrow-band controller drives the RF input power of the AOM and the AM component of beam 1 at $f_m$ is rejected by about 70 dB. However, it is not possible to reach such a rejection on beam 2. The highest level obtained with a very careful alignment of photodiode PD2 on the beam is 50 dB only [2]. It must be noted that this rejection cannot be reached when the splitting between correction beam and useful beam is made at the output of the AOM, before the telescope. Moreover, one observes that any partial occultation of beam 2 with the vertical edge of a screen results in the detection of a large amplitude modulation at $f_m$, although the correction on beam 1 remains excellent.

These effects are due to the angular deviation of the beam in the horizontal plane associated to the frequency shift provided by the AOM. This leads to several beams at the output of the AOM, each of them corresponding to one of the Fourier components applied to its RF input. The angular shift between two successive components is $\lambda f_m / v_{ac}$ where $\lambda$ is the optical wavelength and $v_{ac}$ is the acoustical celerity in the AOM. In order to limit this effect, the AOM was located at the focus of lens L, which transforms the angular deviation of the frequency components into a lateral shift. Nevertheless, the occultation of a part of beam 2 results in the attenuation of some of its Fourier components and gives rise to an amplitude modulation of the beam intensity at $f_m$ that does not exist on beam 1 and cannot be corrected. In particular, because of its finished area, a photodiode may act as a diaphragm on the beam.

If photodiodes PD1 and PD2 are not aligned onto beams 1 and 2 exactly in the same way, an additional AM appears on beam 2 which cannot be cancelled by the controller. Fig. 2 shows the normalized photocurrent component at $f_m$ generated by photodiode PD2 against the relative location of the edge of an occulting screen. Beam 2 propagates in $z$ direction and its Fourier components are split in $x$ direction by the AOM. The beam is partially occulted between $x = -\infty$ to $x = X$ by the screen ($x = 0$ corresponds to the beam centre), whose edge is parallel to $y$ direction. The screen is located close to the beam waist (radius $w_0$). For these measurements, the AM correction system was active and photodiode PD2 was carefully located to obtain the maximum AM rejection on beam 2 without the screen. The continuous line in Fig. 2 is computed from the power received by photodiode PD2 considering that two successive Fourier components are shifted by $A$ in the $x$ direction. In our experimental set-up, $w_0 = 4\,\text{mm}$ and $A = 0.336\,\text{mm}$. These values derived from the AOM parameters and the optical scheme were checked using a beam analyser. A good agreement with the experimental data is observed.

A high level of AM rejection can be reached using the set-up depicted in Fig. 3. The light is coupled into a 50 cm long polarisation maintaining single mode optical fibre located at the output of the AOM which cancels the splitting of the Fourier components of the modulated beam. The beam is then split into the correction beam (beam 1) and the useful beam (beam 2). Only beam 2 is passed through the telescope. Fig. 4 and 5 show the frequency analysis of the photocurrents provided by PD1 and PD2 respectively near the modulation frequency $f_m = 2.5\,\text{MHz}$. Without any correction, a large AM component is observed at 2.5 MHz, as the low-frequency noise components transferred at this carrier frequency. With the correction on, the noise level of beam 2 is practically reduced to the detector noise level. The AM component at 2.5 MHz is rejected by 67 dB. This value, close to the rejection observed on the error signal (Fig. 4), is only 20 dB above the detector noise level. To obtain this result, the λ/2 plate located at the output of the fibre is adjusted in order to optimise the rejection on beam 2 (Fig. 3). As a matter of fact, the beam splitter behaves somewhat as an analyser for the beam polarisation. A $10^{-5}$ residual AM of the useful beam is then achieved. We have checked that the rejection level is independent from the location of PD2 on the beam and from any partial occultation of the beam. It can be asserted that our system really cancels the RAM generated by the AOM.

In conclusion, we have demonstrated in this letter a laser beam FM modulator with a residual AM level of $10^{-5}$ at the modulation frequency. In our system, the frequency modulation is produced by an AOM whose frequency band makes possible modulation frequencies of a few megahertz with a modulation index of the order of unity. At the AOM output, the beam is coupled into a single mode optical fibre to overcome the limitation of the rejection possibility due to the angular shift of the beam Fourier components. The method described here is easy to implement. Moreover, the fibre provides an efficient spatial filtering and gives a $TEM_{00}$ mode at its output.

**Figure captions**

**Fig. 1:**  Set-up for AM rejection.

beam 1: correction beam ;

beam 2: useful beam. It crosses the iodine cell during spectroscopy experiments.

**Fig. 2:**  Normalized photocurrent component at $f_m$ generated by photodiode PD2 against the relative location of the edge of an occulting screen on the beam.

$i^{(f_m)}$ : photocurrent component at $f_m$; $\rho$ : photodiode efficiency; $P_0$ : total power of beam 2 without the screen; $X$: location of the edge of the occulting screen in the $x$ direction (0 corresponds to the beam centre); $w_0$ : radius of the beam waist.

**Fig. 3:**  Set-up for cancellation of the Fourier components splitting of the modulated beam.

**Fig. 4:**  Frequency analysis of PD1 photocurrent near 2.5 MHz (BW = 30 Hz).

**Fig. 5:**  Frequency analysis of PD2 photocurrent near 2.5 MHz (BW = 30 Hz).

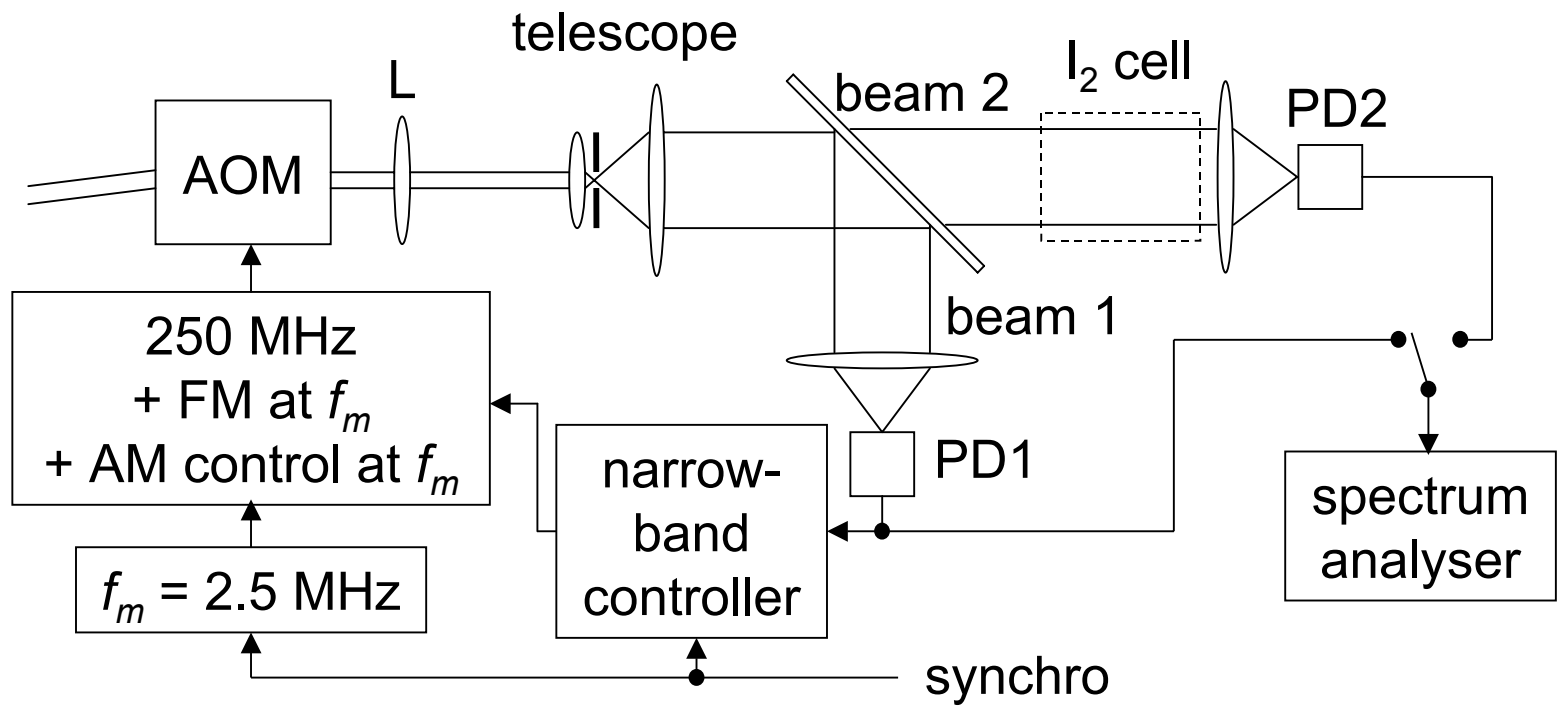

Fig. 1

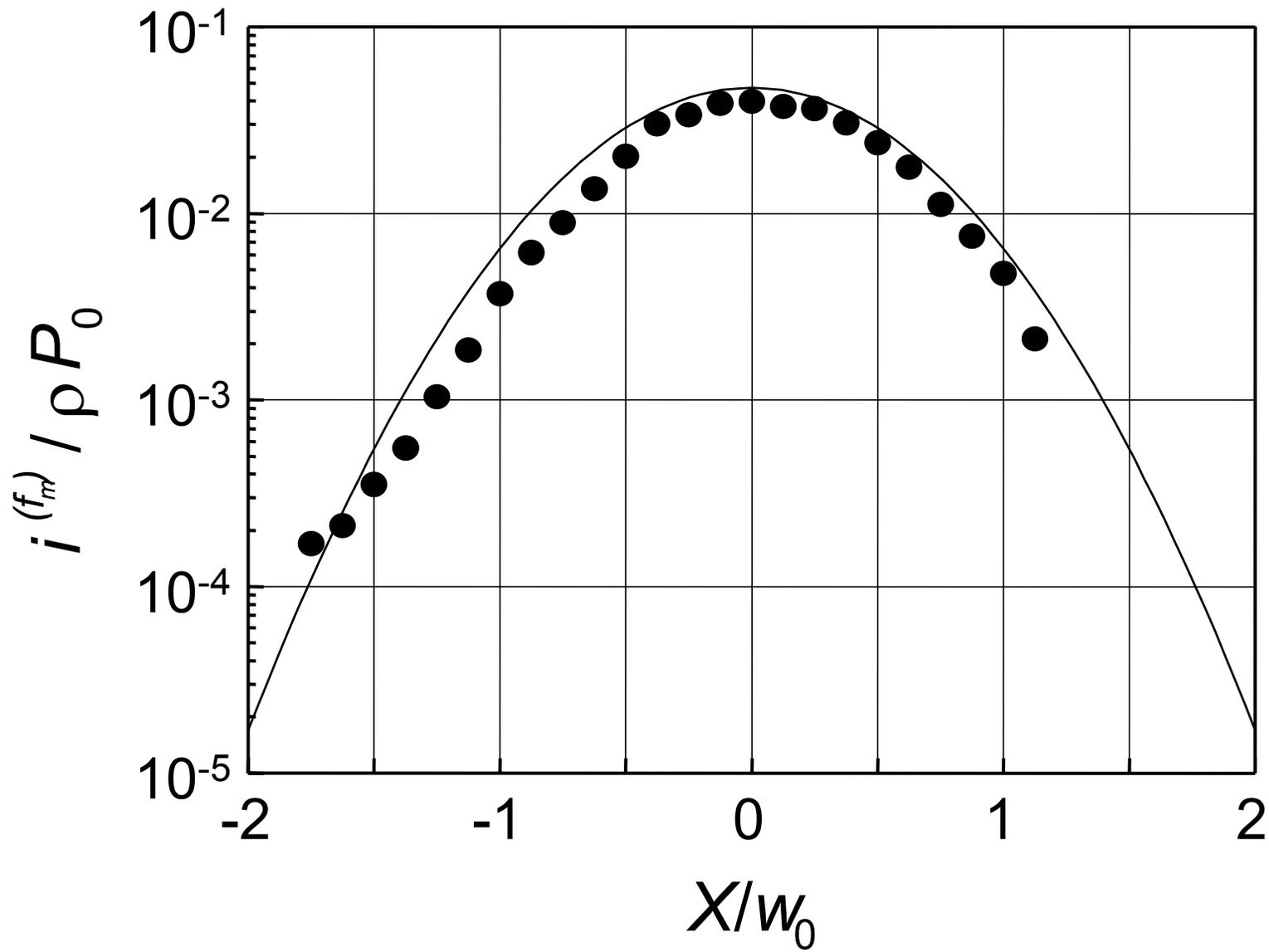

Fig. 2

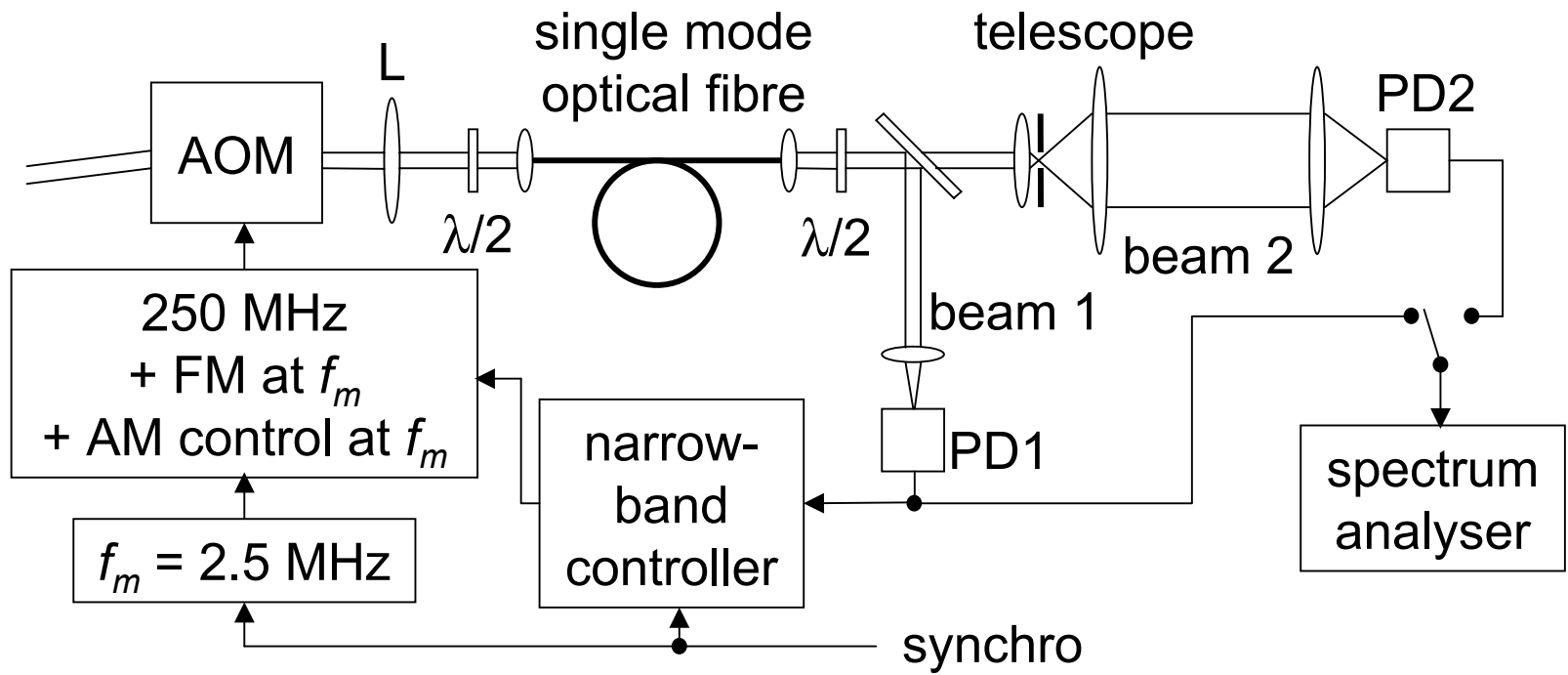

Fig. 3

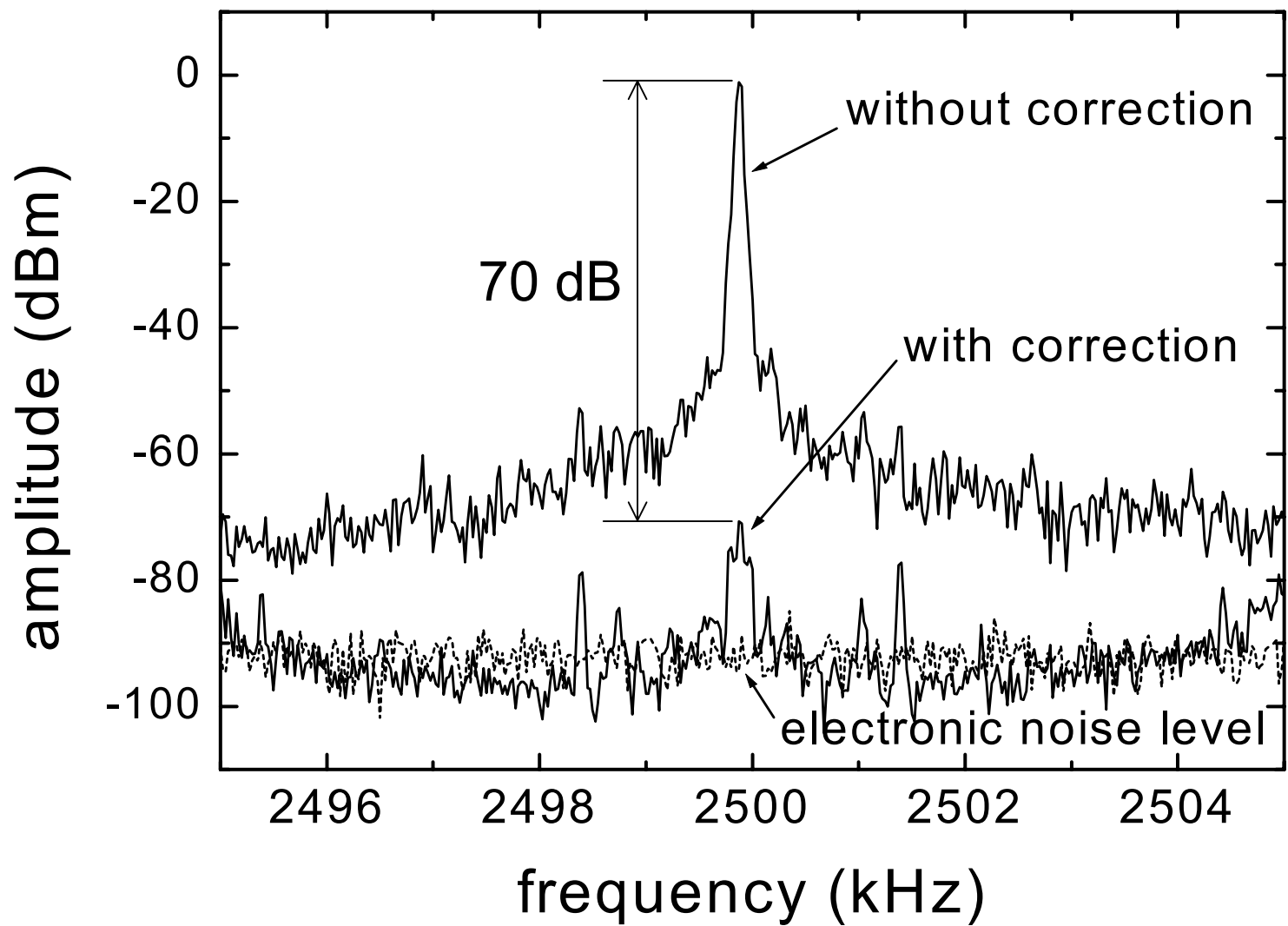

Fig. 4

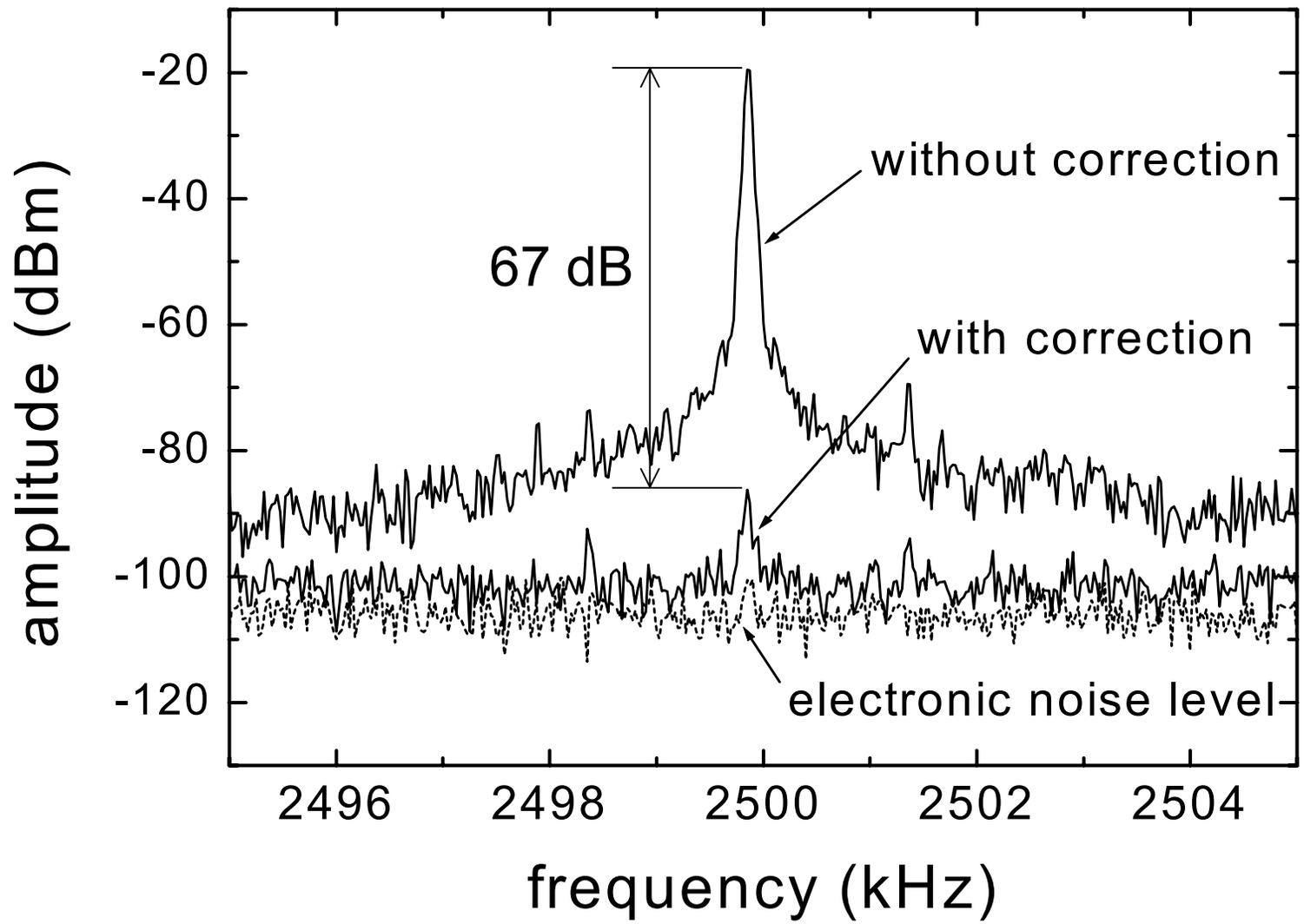

Fig. 5